\documentclass{article}
\begin{document}
\def\lambdabar{{\mathchar'26\mkern-9mu\lambda}}
\baselineskip20pt

\begin{center}
{\large Gravitational Couplings of Intrinsic Spin}\\

\vspace{.25in}

\baselineskip16pt
Bahram Mashhoon\\
Department of Physics and Astronomy\\
University of Missouri-Columbia\\
Columbia, Missouri 65211, USA
\end{center}

\vspace{.5in}

\baselineskip12pt
\begin{center}
Abstract
\end{center}

The gravitational couplings of intrinsic spin are briefly reviewed.  A consequence of the Dirac equation in the exterior
gravitational field of a rotating mass is considered in detail, namely, the difference in the energy of a
spin-$\frac{1}{2}$ particle polarized vertically up and down near the surface of a rotating body is $\hbar\Omega\;{\rm
sin}\;\theta$.  Here
$\theta$ is the latitude and $\Omega = 2GJ/(c^{2}R^{3})$, where $J$ and $R$ are, respectively, the angular
momentum and radius of the body.  It seems that this relativistic quantum gravitational effect could be measurable in the
foreseeable future.  

\newpage
\noindent{\large 1 Introduction}\\

About forty years ago, Kobzarev and Okun [1] considered the theoretical possibility that a nuclear particle may possess a
gravitoelectric dipole moment.  This would lead to a violation of the equivalence principle through an interaction of the
form $H_{{\rm int}}=A\;\mbox{\boldmath$\sigma$}\cdot{\bf g}$, where $A$ is an amplitude, $\mbox{\boldmath$\sigma$}$ is the particle
spin and {\bf g} is the gravitational acceleration due to a massive body such as the Earth.  Similar spin-gravitoelectric
couplings of the form
$f(r)\;\mbox{\boldmath$\sigma$}\cdot{\bf\hat{r}}$ have been considered by a number of authors in connection with the possible
breakdown of parity and time reversal invariance in gravitation [2].  Leitner and Okubo used the hyperfine splitting of the
ground state of hydrogen to set an upper limit on the strength of such an interaction [2].  Meanwhile, Dabbs {\it et al.}
[3] studied the free fall of neutrons polarized vertically up and down in the gravitational field of the Earth and found no
splitting in the gravitational acceleration greater than a few percent of $g$.  However, a few years later observational
evidence was reported for the gravitoelectric dipole moment of the proton [4].  This was soon shown to be spurious by the
experiments of Vasil'ev [5] and Young [6].  In particular, Young [6] placed an upper limit of 0.3 Hz on the gravity shift of
the proton Larmor frequency in 1969.  Finally, a significant upper limit of $10^{-4}$ Hz  was placed on a possible shift
of the deuteron Larmor frequency due to the Earth's gravitational field by Wineland and Ramsey [7] in 1972.  

In 1989, the observation of an anomalous difference in the weight of mechanical gyroscopes rotating vertically upward and
downward was reported [8].  Again, the existence of such a rotor weight change was soon contradicted by subsequent
experiments [9].

The observational search for the role of intrinsic spin in the gravitational interaction as well as the spacetime torsion has
continued and many significant experiments have been performed [10-16].  These experiments have also explored finite-range axionlike
interactions, which could be of the ${\bf\hat {r}}\cdot\mbox{\boldmath$\sigma$}$ (``monopole-dipole'') form as well as a linear
combination of
$\mbox{\boldmath$\sigma$}_A\cdot\mbox{\boldmath$\sigma$}_B$ and ${\bf \hat
{r}}\cdot\mbox{\boldmath$\sigma$}_A\;\;{\bf\hat{r}}\cdot\mbox{\boldmath$\sigma$}_B$ (``dipole-dipole'') form, and have
placed useful restrictions on the parameters of such interactions.  Indeed, the past few decades have witnessed the
emergence of extremely precise measurement techniques [17] that make it possible to detect frequency shifts of order
$10^{-9}$ Hz, an improvement of five orders of magnitude over what was possible three decades ago [7].

The aim of the present paper is to discuss the gravitomagnetic coupling of intrinsic spin due to the fact that according
to the standard theory a spinning particle possesses a gravitomagnetic dipole moment.  This moment couples to the
gravitomagnetic field of a rotating mass (such as the Earth) in complete analogy with the
$-\mbox{\boldmath$\mu$}\cdot{\bf B}$ interaction in electrodynamics.  Instead of treating the Dirac equation in the
exterior gravitational field of a rotating mass, a heuristic derivation of this general interaction is given in sections 2
and 3 on the basis of the gravitational Larmor theorem.  For a spin-$\frac{1}{2}$ particle near the surface of the Earth,
the effect involves a frequency shift of order $10^{-14}$ Hz. Section 4 contains a brief discussion of the prospects for
the measurement of this relativistic quantum gravitational effect. 

\vspace{.25in}
\noindent {\large 2 Inertia of Intrinsic Spin}\\

Imagine an observer in a laboratory on the Earth using Earth-based coordinate axes to describe the results of
measurements.  The particles involved in the experiments on the rotating Earth are waves propagating in inertial spacetime
and it is natural to assume that they would keep their polarization aspects fixed in the underlying inertial frame.  As
measured by the observer, however, such intrinsic spin must ``precess'' in a sense opposite to the sense of rotation of
the Earth.  The Hamiltonian associated with such motion would be of the form
$H = -\mbox{\boldmath$\sigma$}\cdot\mbox{\boldmath$\Omega$}$, where $\mbox{\boldmath$\Omega$}$ is the frequency of rotation
of the laboratory frame.  The existence of such a Hamiltonian would show that intrinsic spin has rotational inertia.  In
quantum mechanics, mass and spin characterize the irreducible unitary representations of the inhomogeneous Lorentz group.
The inertial properties of mass are well known in classical mechanics through various translational and rotational
acceleration effects.  In quantum mechanics, the inertial properties of mass have been experimentally investigated by a
number of authors [18].  It is therefore interesting to consider the inertial properties of spin [19].

The coupling of intrinsic spin with rotation indicated above may be illustrated by a simple example.  Imagine an observer
rotating counterclockwise with uniform frequency $\Omega$ about the direction of propagation of a plane linearly polarized
monochromatic electromagnetic wave of frequency $\omega \gg \Omega$.  For instance, the observer could be in an
Earth-based laboratory and
$\Omega$ would then be the frequency of the proper rotation of the Earth.  We neglect gravitational effects in this section
and consider all phenomena in a global inertial frame in Minkowski spacetime.  Let the observer move on a circle of radius
$r$ with speed $c \beta = r \Omega$ in the $(x,y)-$ plane of the inertial frame and let the electric field of the wave be
given by the real part of 

\begin{equation}
{\rm {\bf E}} = E_{0} \: {\bf \hat{x}} \: e^{-i \omega t + ikz},
\end{equation}

\noindent where $E_{0}$ is a constant amplitude, ${\bf k} = k \; {\bf \hat z}$ is the wave vector and $\omega = ck$. 
>From the viewpoint of the rotating observer, the direction of linear polarization that is fixed in the inertial frame must
drift in a clockwise sense about the direction of propagation, i.e.

\begin{equation}
{\rm {\bf E}}=E_{0} \left ({\rm cos}\; \Omega t \: {\bf
\hat{x}}^{\prime}-{\rm sin} \; \Omega t \: {\bf \hat{y}}^{\prime}
\right ) e^{-i
\omega t + ikz},
\end{equation}

\noindent where ${\bf \hat{x}}^{\prime}={\bf \hat{x}} \; {\rm cos} \; \Omega t + {\bf \hat{y}} \; {\rm sin} \; \Omega t, {\bf
\hat{y}}^{\prime}= -{\bf \hat{x}} \; {\rm sin} \;
\Omega t + {\bf \hat{y}} \; {\rm cos} \; \Omega t$ and ${\bf \hat{z}}^{\prime}={\bf
\hat{z}}$ denote the Cartesian coordinate axes in the rotating frame of the observer.  Specifically, the two coordinate
systems are related by a simple rotation such that

\begin{equation} {\bf \hat{x}}+i {\bf \hat{y}}=e^{\pm i \Omega t} \left ( {\bf
\hat{x}}^{\prime} \pm i {\bf \hat{y}}^{\prime}
\right ).
\end{equation}

\noindent The linearly polarized wave (1) is a coherent superposition of a right circularly polarized (RCP) wave and
a left circularly polarized (LCP) wave, i.e.

\begin{equation}
{\rm {\bf E}}=\frac{1}{2}E_{0} \left ({\bf \hat{x}}+i{\bf \hat{y}} \right ) e^{-i \omega t+ikz}+\frac{1}{2}E_{0}
\left ({\bf \hat{x}}-i{\bf \hat{y}}
\right ) e^{-i \omega t+ikz}.
\end{equation}

\noindent From the viewpoint of the rotating observer, these eigenstates of the radiation field remain invariant,

\begin{equation}
{\rm {\bf E}} = \frac{1}{2} E_{0} \left ( {\bf \hat{x}}^{\prime}+i {\bf
\hat{y}}^{\prime} \right ) e^{-i(\omega-\Omega) t+ikz}+\frac{1}{2} E_{0} \left ({\bf \hat{x}}^{\prime}-i{\bf
\hat{y}}^{\prime} \right ) e^{-i(\omega+\Omega)t+ikz},
\end{equation}

\noindent except that the frequency of the RCP component is perceived to be
$\omega-\Omega$ while that of the LCP wave is perceived to be $\omega + \Omega$ with respect to inertial time $t$.  The
proper time of the observer is, however,
$\tau = t/\gamma$, where $\gamma = \left (1-\beta^{2} \right ) ^{-1/2}$.  Thus we find that the proper frequencies
measured by the observer are\

\begin{equation}
\omega^{\prime}=\gamma(\omega \mp \Omega).
\end{equation}

\noindent Here the Lorentz factor accounts for time dilation, which is all that should happen according to the
transverse Doppler effect.  Instead, we have in (6) the additional ``angular Doppler terms'' $\mp \Omega$ that have the
following physical origin:  In an RCP (LCP) wave, the electric and magnetic fields rotate in the positive (negative) sense
about the direction of propagation with frequency $\omega$.  Since the observer rotates in the positive sense with
frequency $\Omega$, it perceives the effective frequency of the RCP (LCP) wave to be $\omega - \Omega \; (\omega +
\Omega)$.  In the JWKB limit,
$\omega \rightarrow \infty$ and the ``angular Doppler terms'' disappear since $\mp
\Omega/\omega \rightarrow 0$.  Our heuristic treatment ignores certain relativistic corrections that are not essential for the
purposes of this discussion.  Writing equation (6) in terms of the photon energy as
$E^{\prime}=\gamma(E\mp
\hbar \Omega)$, we see that the deviation from the simple transverse Doppler effect is due to the coupling of the spin of
a circularly polarized photon to the rotation of the observer, since a RCP (LCP) photon carries an intrinsic spin of
$\hbar (- \hbar)$ along its direction of propagation [20].  These elementary considerations already contain the basic
aspects of the phenomenon of spin-rotation coupling, as can be seen from the following discussion based on the theory of
relativity [21].

The special theory of relativity consists of two main elements:  the principle of relativity (i.e. Lorentz invariance) and the
hypothesis of locality.  The latter specifies what an accelerated observer measures by establishing a connection between the
accelerated observer and an inertial observer.  Indeed, it requires that an accelerated observer be at each instant locally equivalent
to a momentarily comoving inertial observer.  This is a nontrivial axiom since there exist definite acceleration scales of
time and length that are associated with an accelerated observer.  In the case under consideration, e.g., the acceleration
length of the rotating observer is ${\cal L} = c / \Omega$ and the corresponding temporal scale is ${\cal L}/ c =
\Omega^{-1}$.  Moreover, an elementary application of the hypothesis of locality would imply that
$\omega^{\prime}=\gamma \omega$ by the transverse Doppler effect, since the connection between the instantaneous inertial
frame of the accelerated observer and our global inertial frame simply results in the standard Doppler and aberration
formulas with a time-dependent velocity $c\mbox{\boldmath $\beta$} (t)$.  On the other hand, it should be clear that to
measure wave characteristics such as the frequency, one must observe at least a few periods of the oscillations of the
wave before a determination of the frequency becomes even possible.  In this way, the curvature of the observer's
worldline would have to be taken into consideration, and hence the standard Doppler and aberration formulas of
relativity theory are valid only to the extent that the period of the wave ${T}=2 \pi/\omega$ is negligible compared to
$\Omega^{-1}$, i.e. $\Omega T=2 \pi \Omega/\omega \rightarrow 0$.  In view of these remarks, it is therefore natural to
apply the locality axiom only to the electromagnetic field; then, the measured field could be Fourier analyzed -- which is
a nonlocal operation -- to obtain the frequency and wave vector content of the field in the accelerated frame.  This is
indeed the physical basis for the extension of relativistic wave equations to accelerated systems; in fact, this extended
hypothesis of locality for wave phenomena is equivalent to the assumption of minimal coupling.  Using this approach, one
finds that for $\omega \gg \Omega$, the standard Doppler and aberration formulas should be modified to

\begin{equation}
\omega^{\prime}=\gamma (\omega-c \mbox{\boldmath $\beta$} \cdot {\rm {\bf k}}) - \gamma {\bf \hat{H}} \cdot
\mbox{\boldmath
$\Omega$},
\end{equation}

\begin{equation}
{\rm {\bf k}}^{\prime}= {\rm {\bf k}} + (\gamma-1) (\hat{\mbox{\boldmath $\beta$}} \cdot {\rm {\bf k}})
\hat{\mbox{\boldmath $\beta$}} -
\frac{1}{c} \gamma \omega \mbox{\boldmath $\beta$} + \frac{1}{c} \gamma ( {\bf
\hat{H}} \cdot \mbox{\boldmath $\Omega$})
\mbox{\boldmath
$\beta$},
\end{equation}

\noindent where ${\bf \hat{H}} = \pm c {\rm {\bf k}}/ \omega$ is the unit helicity vector.  One can then
consider optical interferometry in a rotating frame that would be based on the spin of the photon in contrast to the
Sagnac effect that is connected to its orbital angular momentum [21].

The general expression for spin-rotation coupling can be written as

\begin{equation}
E^{\prime}=\gamma (E-\hbar M \Omega),
\end{equation}

\noindent where $M$ is the total (orbital plus spin) ``magnetic'' quantum number along the axis of rotation; that is,
$M=0, \pm 1, \pm 2, \ldots$ for a scalar or a vector field while $M \mp
\frac{1}{2}=0, \pm1, \pm2, \ldots$ for a Dirac field.  In the JWKB approximation, equation (9) can be written as
$E^{\prime}=\gamma(E- \mbox{\boldmath $\Omega$} \cdot {\rm {\bf J}})$, where ${\rm {\bf J}}={\rm {\bf L}}+{\rm {\bf
S}}={\rm {\bf r}} \times {\rm {\bf P}} +
{\rm {\bf S}}$.  Thus
$E^{\prime}=\gamma(E- {\rm {\bf v}}\cdot {\bf P})-\gamma
{\rm {\bf S}}\cdot \mbox{\boldmath
$\Omega$}$, so that in the absence of intrinsic spin we recover the classical expression for the energy of a particle as
measured in the rotating frame with
${\rm {\bf v}}= \mbox{\boldmath $\Omega$} \times {\rm {\bf r}}$.  The energy and momentum of a spinning
particle as measured by an accelerated observer are then

\begin{equation}
E^{\prime} = \gamma(E - {\rm {\bf v}}\cdot {\rm {\bf P}}-{\rm {\bf S}}\cdot\mbox{\boldmath$\Omega$})\;\;,
\end{equation}

\begin{equation} 
{\rm {\bf P}}^{\prime} = {\rm {\bf P}} + (\gamma -
1)\;({\rm {\bf P}}\cdot\hat{\mbox{\boldmath${\beta}$}})\hat{\mbox{\boldmath${\beta}$}} - \frac{1}{c}\gamma
E\mbox{\boldmath$\beta$} + \frac{1}{c}\gamma({\rm {\bf S}}\cdot\mbox{\boldmath$\Omega$})\mbox{\boldmath$\beta$}\;\;,
\end{equation}

\noindent using the same JWKB approach as in the derivation of equations (7) - (8).  It follows that

\begin{equation}
E^{\prime 2} - c^{2}P^{\prime 2} = m^2c^4 - 2E(\mbox{\rm {\bf S}}\cdot\mbox{\boldmath$\Omega$}) +
({\rm {\bf S}}\cdot\mbox{\boldmath$\Omega$})^2\;\;.
\end{equation}

\noindent These results reduce to the equations appropriate for light once we set $E=\hbar\omega, {\rm {\bf P}} =
\hbar{\rm {\bf k}}, {\rm {\bf S}} = \hbar\hat{{\rm {\bf H}}}$ and $m=0$.

Experimental evidence for helicity-rotation coupling exists in the microwave and optical regimes via the phenomenon of
frequency shift of polarized radiation [19].  Moreover, there is observational evidence for the coupling of spin-$\frac{1}{2}$
particles with the rotation of the Earth [19].  The analogous gravitational coupling of intrinsic spin is considered in the
next section. 

\newpage

\noindent{\large 3 Spin-Gravitomagnetic Coupling}\\

To extend the physics of spin-rotation coupling to the gravitational field, one must resort to Einstein's heuristic
principle of equivalence.  It is possible to interpret this principle in the post-Newtonian approximation via the
gravitational Larmor theorem [22].  Newton's law of gravitation is formally analogous to Coulomb's law of electricity;
therefore, one may describe Newtonian gravitational effects in terms of a gravitoelectric field.  The classical tests of
general relativity are all due to post-Newtonian gravitoelectric corrections.  However, any consistent framework that brings
Newtonian gravitation and Lorentz invariance together must of necessity contain a gravitomagnetic field that would be due
to mass current.  A direct measurement of the gravitomagnetic field of the Earth via the precession of superconducting
gyroscopes in a polar orbit about the Earth is one of the goals of the Stanford gyroscope experiment (GP-B) planned for
2001.

In the linear approximation of general relativity, where gravitational effects are treated as linear perturbations in a
global inertial frame in Minkowski spacetime, one can express the gravitational field equations as Maxwell's equations for
the gravitoelectric field ${\rm {\bf E}}_{g}$ and the gravitomagnetic field ${\rm {\bf B}}_{g}$ once $O(c^{-4})$
terms are neglected in the post-Newtonian metric perturbations [22].  Specifically, we let $g_{\mu\nu}=\eta_{\mu\nu} +
h_{\mu\nu}$, where $\eta_{\mu\nu}$ is the Minkowski metric and for the linear perturbation $h_{\mu\nu}$ we define
$\bar{h}_{\mu\nu} = h_{\mu\nu} - \frac{1}{2}\eta_{\mu\nu}h^{\alpha}_{\alpha}$.  Then
$\bar{h}^{00}=4\phi_{g}/c^2\;,\; \bar{h}^{0 i} = 2 A^i_g/c^2$ and $\bar{h}^{ij} = O(c^{-4})$. Here $\phi_g(t,
{\rm {\bf x}})$ is the gravitoelectric potential and ${\rm {\bf A}}_g(t,{\rm {\bf x}})$ is the
gravitomagnetic vector potential.  That is, of the ten effective gravitational potentials $\bar{h}_{\mu\nu}$ in general
relativity, we neglect the six spatial potentials $\bar{h}_{ij}$ as these are of $O(c^{-4})$ for nonrelativistic
(astronomical) sources and from the remaining four potentials one can construct a consistent theory of
gravitoelectromagnetism (GEM) in this approximation scheme.  Let us note that
$\bar{h}^{0\mu}=2c^{-2}(2\phi_{g}\;,\;{\rm {\bf A}}_{g})$, so that the Lorentz gauge condition
$\bar{h}^{\mu\nu}\!_{,\nu}=0$ reduces in this case to 

\begin{equation}
\frac{2}{c}\;\;\frac{\partial\phi_{g}}{\partial t} + \mbox{\boldmath$\nabla$}\cdot{\rm {\bf A}}_g = 0\;\;.
\end{equation}

\noindent Thus $A^{\mu} = (2\phi_{g}\;,\;{\rm {\bf A}}_g)$ is the effective GEM potential and the spacetime metric is
given by 

\begin{equation}
ds^2 = -c^2\left (1-\frac{2\phi_g}{c^2}\right )dt^2 - \frac{4}{c}\left ({\bf A}_g\cdot d{\bf x}\right ) dt + \left (1
+ \frac{2\phi_{g}}{c^2}\right ) \delta_{ij} dx^idx^j\;\;.
\end{equation}

\noindent The analogy with electrodynamics turns out to be exact, except for the fact that the ratio of the gravitomagnetic
charge to the gravitoelectric charge is two, $q_B/q_E = 2$; that is, linear gravity is a spin-2 field in contrast to the
spin-1 character of the electromagnetic field that implies $q_B/q_E = 1$ for the Maxwell theory.

In electrodynamics, Larmor established a theorem regarding the local equivalence of magnetism and rotation for all charged
particles with the same charge-to-mass ratio $q/m$.  In fact, the electromagnetic field can be locally replaced by an
accelerated frame with translational acceleration ${\rm {\bf a}}_{L} = -(q/m){\rm {\bf E}}$ and rotational
(Larmor) frequency $\mbox{\boldmath$\omega$}_{L} = q{\rm {\bf B}}/(2mc)$.  In electrodynamics, $q/m$ can be positive,
zero or negative; however, the gravitational charge-to-mass ratio is universal due to the experimentally well-tested
equivalence of gravitational and inertial masses.  This leads directly to Einstein's principle of equivalence and hence a
geometric theory of gravitation.  Einstein's heuristic principle of equivalence traditionally involves the local
equivalence of the gravitoelectric field with the translational acceleration of the ``Einstein elevator'' in Minkowski
spacetime.  The interpretation of Einstein's principle in terms of the gravitational Larmor theorem would then involve, in
addition, the local equivalence of the gravitomagnetic field with the Larmor rotation of the elevator as well.

Let us consider the exterior field of an almost spherical rotating astronomical body (such as the Earth) with GEM
potentials

\begin{equation}
\phi_g \simeq \frac{GM}{r}\;\;,\;\;{\rm {\bf A}}_g \simeq \frac{G}{c}\;\frac{{\rm {\bf J}} \times
{\rm {\bf r}}}{r^3}\;\;,
\end{equation}

\noindent where $M$ is the mass and $J$ is the angular momentum of the source.  These potentials can be obtained from the
electromagnetic analogy by assuming that the source has positive gravitoelectric charge $Q_E=M$ and gravitomagnetic charge
$Q_B=2M$.  The GEM fields are then 

\begin{equation}
{\rm {\bf E}}_g = -\mbox{\boldmath$\nabla$}\phi_g - \frac{1}{2c}\;\frac{\partial}{\partial
t}{\rm {\bf A}}_g\;\;,\;\;{\rm {\bf B}}_g = \mbox{\boldmath$\nabla$}\times{\rm {\bf A}}_g\;\;.
\end{equation}

\noindent The motion of test particles in the gravitational field of a rotating mass can be obtained from the Lorentz force
law if we assume that for a test particle of inertial mass $m$ the gravitational charges are negative, i.e.
$q_E = -m$ and $q_B = -2m$, in order to take due account of the dominant gravitational attraction between the test particle
and the source.  It turns out that an ideal test gyroscope at rest outside the rotating source undergoes gravitomagnetic
precession

\begin{equation}
\frac{d{\rm {\bf S}}}{dt} = \mbox{\boldmath$\Omega$}_{P} \times{\rm {\bf S}}
\end{equation}

\noindent with frequency

\begin{equation}
\mbox{\boldmath$\Omega$}_P = \frac{1}{c}{\bf {B}}_g =
\frac{GJ}{c^2r^3}\left[3\left({\bf \hat{r}}\cdot{\bf\hat{J}}\right){\bf\hat{r}} -
{\bf\hat{J}}\right]\;\;.
\end{equation}

\noindent Imagine now that we replace the gravitomagnetic field by a rotating frame in the neighborhood of the gyroscope. 
As referred to observers at rest in the rotating frame, the motion of the gyroscope would be the same as before if the
observers rotate with Larmor frequency $\mbox{\boldmath$\omega$}_L = -\mbox{\boldmath$\Omega$}_P$.  This relation is
consistent with the Larmor formula $\mbox{\boldmath$\omega$}_L = q{\bf B}/(2mc)$ once we set $q_B = -2m$ and ${\bf B}_g =
c\mbox{\boldmath$\Omega$}_P$ as in equation (18).  Thus a consistent and complete gravitoelectromagnetic formalism can be
developed along these lines [22].  

In particular, the spin-rotation coupling can be extended to gravitomagnetism via the Larmor theorem with
$\mbox{\boldmath$\omega$}_L = -\mbox{\boldmath$\Omega$}_P$.  That is, the interaction of intrinsic spin with the
gravitomagnetic field is given by the Hamiltonian $H=\mbox{\boldmath$\sigma$}\cdot\mbox{\boldmath$\Omega$}_P$, since this
interaction in the Larmor frame would be $H= -\mbox{\boldmath$\sigma$}\cdot\mbox{\boldmath$\omega$}_L$ as described in
section 2.  Moreover, the Heisenberg equations of motion for the spin-gravity interaction
$H=\mbox{\boldmath$\sigma$}\cdot\mbox{\boldmath$\Omega$}_P$ are formally the same as equations (17) - (18) for the
precession of an ideal test gyroscope. 

In classical electrodynamics, the magnetic dipole moment for a particle of mass
$m$ and charge $q$ is given by $\mbox{\boldmath$\mu$}=q{\bf S}/(2mc)$, where ${\bf S}$ is its orbital angular momentum. The
energy associated with the interaction of this magnetic moment with a magnetic field ${\bf B}$ is
$-\mbox{\boldmath$\mu$}\cdot{\bf B}$.  Extending these notions to GEM with $q_B = -2m$, we find that a gravitomagnetic
dipole moment for a gyroscope of spin ${\bf S}$ is $\mbox{\boldmath$\mu$}_g = -{\bf S}/c$ and the energy of interaction
with a gravitomagnetic field is $-\mbox{\boldmath$\mu$}_g\cdot{\bf B}_g = {\bf S}\cdot\mbox{\boldmath$\Omega$}_P$.  A
further extension of this result to the intrinsic spin of particles naturally leads to the interaction Hamiltonian
$H=\mbox{\boldmath$\sigma$}\cdot\mbox{\boldmath$\Omega$}_P$.  The gravitoelectric analog of this interaction has already been
discussed in section 1; that is, $H_{\rm int}=-{\bf d}_{g} \cdot {\bf E}_{g}$, where ${\bf d}_{g}=A \mbox{\boldmath$\sigma$}$ would
be the hypothetical gravitoelectric dipole moment of a particle [1,2] and ${\bf E}_{g}= -{\bf g}$ from (15) and (16).

Let us imagine an experiment in a laboratory near the surface of an astronomical body (such as the Earth) involving the
difference in the energy of a particle of spin $\sigma = s\hbar$ polarized vertically up and down (i.e. perpendicular to
the surface).  According to the spin-gravitomagnetic coupling, the result is 

\begin{equation}
E_{+}-E_{-} = 2s\hbar\Omega \; {\rm sin} \; \theta\;\;.
\end{equation}

\noindent Here $\theta$ is the geographic latitude (i.e. $E_{+}=E_{-}$ at the equator) and
$\Omega$ is the effective frequency associated with the gravitomagnetic field

\begin{equation}
\Omega = \frac{2GJ}{c^2R^3}\;\;,
\end{equation}

\noindent where $R$ is the mean radius of the body.  Equation (19) expresses a relativistic quantum gravitational effect;
indeed, one can write $\hbar\Omega=(2cJ/R^3)L^2_P$, where $L_P = (\hbar G/c^3)^{1/2}$ is the Planck length.  Let us note
that for the Earth $\hbar\Omega_E\simeq 2\times 10^{-29}{\rm eV}$, while near the surface of Jupiter $\hbar\Omega_J\simeq
10^{-27}{\rm eV}$; similarly, for the Sun $\hbar\Omega_S \sim 10^{-27}{\rm eV}$, but for a neutron star $\hbar\Omega_{NS}\sim
10^{-14}{\rm eV}$.

It is important to point out that the spin-rotation-gravity coupling has appeared in the work of many authors who have
studied wave equations in accelerated systems and gravitational fields [23,24].  In particular, the $\mbox{\boldmath$\sigma$}
\cdot \mbox{\boldmath$\Omega$}_P$ interaction under scrutiny in this work first appeared in the work of de Oliveira and Tiomno
[23].  The observation of wave phenomena associated with such couplings was first independently investigated in connection with
possible limitations of the general theory of relativity in [25].  Dynamics in electromagnetic fields can be generated by the
transformation of the momentum via $p_{\mu}\rightarrow p_{\mu} - (q/c)A_{\mu}$, where $A_{\mu}=(-\phi, {\bf A})$ is the EM
potential.  The same holds in the GEM case, except that the analog of $A_{\mu}$ is $(-2\phi_{g}, {\bf A}_g)$.  Let us consider, for
instance, the motion of electromagnetic waves in the exterior field of a rotating mass.  The effective gravitational charge in this
case should be determined based on the fact that a photon of energy $\hbar\omega$ in ``cyclotron'' motion has an effective inertial
mass of $\hbar\omega/c^2$ and hence the effective GEM charges are $q_{E}= -\hbar\omega/c^2$ and $q_{B}= -2\hbar\omega/c^2$.  The
eigenvalue problem in gravitomagnetic fields leads to discreteness properties for the modes reminiscent of the Fock-Darwin-Landau
levels in a magnetic field.  Imagine, for instance, the motion of electromagnetic waves in a gravitomagnetic field characterized by
the magnitude of the effective ``cyclotron'' frequency $\Omega_c\simeq 2GJ/(c^2r^3)$.  It follows from the explicit solution of
Maxwell's equations in this background [25] that the wave functions are proportional to Hermite polynomials.  These polynomials vary
over a harmonic characteristic lengthscale $l_g$ that is given by

\begin{equation}
l_g = \frac{c}{(\omega\Omega_c)^{1/2}}
\end{equation}

\noindent for an electromagnetic mode of frequency $\omega$. If in this equation we set
$\hbar\omega = mc^2$ and $\Omega_c$ = cyclotron frequency in a magnetic field, we recover the {\it magnetic length} that
is well known in the discussion of the motion of a charged particle of mass $m$ in a magnetic field.  It is interesting to
note that the gravitomagnetic acceleration length is given by ${\cal {L}}_g = c/\Omega_c$, so that the {\it gravitomagnetic
length} (21) is the geometric mean of the reduced wavelength of radiation $\lambdabar$ and ${\cal {L}}_g$.  The
gravitomagnetic length $l_g$ is essentially the same as the radius of the ``cyclotron'' orbit for a mode with frequency
equal to the ``cyclotron'' frequency ($\omega =\Omega_c$).  The eigenvalue spectrum clearly shows the existence of a
gravitomagnetic coupling between the photon spin and the rotation of the source [25].  One can show that in the eikonal approximation
the gravitational helicity-rotation coupling leads to a differential deflection of polarized radiation thus violating the
universality of free fall in a gravitational field beyond the geometric optics limit [25,22].  That the spin-gravity interaction
violates the universality of free fall is already apparent from $H=\mbox{\boldmath$\sigma$}\cdot\mbox{\boldmath$\Omega$}_P$, since
this Hamiltonian depends only on the spin of the particle and is independent of its mass.

Imagine, for instance, the scattering of electromagnetic radiation by a black hole (i.e. pure geometry free of matter). 
For a Schwarzschild black hole, the scattering amplitude is independent of the polarization of the incident radiation,
hence the polarization properties of the radiation are preserved in the scattering process.  For a Kerr black hole,
however, the scattering amplitude is dependent upon the polarization of the incident radiation.  It is possible to give
only rough and partial estimates for the motion of {\it wave packets} in a gravitomagnetic field [25,26].  The influence of
helicity-rotation coupling on the gravitational deflection of electromagnetic radiation is rather weak and far below the existing
observational upper limits [27], but could become important in future microlensing experiments with polarized radiation.  To provide
useful astrophysical estimates of the resulting polarization-dependent deflection of radiation, an {\it eikonal} approach has
been developed for the motion of {\it rays} based on equations (7) and (8), i.e. $\omega({\bf {r}}, {\bf {k}}) = ck \pm
{\bf\hat{k}}\cdot \mbox{\boldmath$\Omega$}_P({\bf {r}})$, so that the Einstein deflection is ignored for the sake of
simplicity and only the helicity-rotation coupling is taken into account [22].  In this treatment, the total differential
deflection of positive and negative helicity rays approaching the source together from asymptotic infinity and traveling
to infinity after deflection vanishes in contrast to what is expected from the wave treatment; however, it is possible to
obtain useful estimates for radiation originating near the source.  For instance, consider radiation originating over a
pole and propagating normal to the rotation axis with an impact parameter $D$; then, RCP and LCP waves separate by a total
angle of $\delta \approx 4\lambdabar GJ/(c^{3}D^{3})$ about the average Einstein deflection angle.  A {\it qualitative}
description of this effect is given in [28].  The gravitomagnetic splitting $\delta$ is small; it amounts to about one
milliarcsecond for GHz radio waves passing over a pole of a neutron star.  In addition to this splitting, one expects a
wavelength-independent gravitomagnetic rotation of the plane of polarization along a ray, i.e. the Skrotskii effect that is the
gravitational analog of the Faraday effect [25,29].  Moreover, the difference in the arrival times of positive and negative helicity
radiation originating near a rotating mass and propagating freely outward to a distant point is estimated to be
$T_{+}-T_{-} = - 2\lambdabar G {\bf {J}}\cdot {\bf {r}}/(c^{4}r^{3})$, where ${\bf r}$ is the position vector of the point of origin
of the radiation relative to the center of the rotating source.  This differential time delay due to the different phase speeds of
RCP and LCP waves is too small to be measurable at present [22].

The violation of the universality of free fall is a wave effect, so that it vanishes in the
$\lambdabar/{\cal{L}}_g\rightarrow 0$ limit.  Consider, for instance, a spinning particle in a gravitomagnetic field with
the interaction Hamiltonian $H=\mbox{\boldmath$\sigma$}\cdot\mbox{\boldmath$\Omega$}_P$.  This potential energy is position
dependent; therefore, there exists a gravitomagnetic Stern-Gerlach force ${\bf {F}} = -\mbox{\boldmath$\nabla$}H$ acting
on the particle that is independent of mass and hence violates the universality of the gravitational acceleration. 
Specifically,

\begin{equation}
{\bf F} = \frac{3GJ}{c^{2}r^{4}}
\left \{ [5(\mbox{\boldmath$\sigma$}\cdot{\bf\hat{r}})({\bf\hat{J}}\cdot{\bf\hat{r}})-\mbox{\boldmath$\sigma$}\cdot{\bf\hat
{J}}]{\bf\hat{r}}-
(\mbox{\boldmath$\sigma$}\cdot{\bf\hat{r}}){\bf\hat{J}}-({\bf\hat{J}}\cdot{\bf\hat{r}})\mbox{\boldmath$\sigma$} \right\} \;\;,
\end{equation}

\noindent so that the weight operator for the particle $W=mg-{\bf F}\cdot{\bf\hat{r}}$ is given by $W=mg-3H/r$.  If the
spin is polarized vertically up or down in a laboratory near the Earth,

\begin{equation}
W_{\pm} = mg \mp \frac{3s}{R}\hbar\Omega \; {\rm sin}\; \theta\;\;,
\end{equation}

\noindent so that $W_{\pm} = mg(1\mp\epsilon$), where $\epsilon$ can be expressed as 

\begin{equation}
\epsilon = 6s\left(\frac{I}{MR^2}\right)\;\;\left(\frac{\hbar\omega}{mc^2}\right){\rm sin}\; \theta\;\;.
\end{equation}

\noindent Here $J=I\omega$, $I$ is the moment of inertia and $\omega$ is the proper rotation frequency of the Earth. For a neutron
near the Earth's surface,
$\hbar\omega/(m_nc^2)\simeq 5\times 10^{-29}$; hence, $\epsilon$ is too small to be measurable in the foreseeable future.  It follows
that for polarized materials the relevant $\epsilon$ is expected to be even smaller.  Let us note that $\epsilon$ is directly
proportional to
$\hbar\omega/(mc^2)$, which can be expressed as the ratio of the Compton wavelength of the particle ($\hbar/mc$) to the
rotational acceleration length of the observer ($c/\omega$).  Indeed, the extended nature of the particle makes it
possible for its intrinsic spin to couple to the spacetime curvature resulting in the force ${\bf F}$ that has an exact
analog in the classical Mathisson-Papapetrou spin-curvature force [22,28].  

We have thus far discussed the gravitomagnetic spin-rotation coupling in terms of a single rotating source such as the
Earth.  However, the universality of the gravitational interaction implies that the whole mass-energy content of the
universe is involved in every physical experiment via the gravitational interaction.  In classical physics, the
gravitational force of the rest of the universe enters only through its gradients, which turn out to be rather small for
experiments in the solar system.  The situation is in general different in quantum physics, however.  For instance, in
the calculation of the spin-gravity coupling, the gravitomagnetic field generated by the total mass-energy current must be
taken into account.  This is a difficult problem; however, to get some idea of what is involved here we may use the linear
approximation to write the interaction Hamiltonian as 

\begin{equation}
H=\frac{G}{c^{2}}\;\sum_a\;\frac{3({\bf r}_a\cdot{\bf J}_a)({\bf r}_a\cdot\mbox{\boldmath$\sigma$})-r^2_a({\bf
J}_a\cdot\mbox{\boldmath$\sigma$})}{r^5_a}\;\;,
\end{equation}

\noindent where the sum is over all astronomical sources and ${\bf r}_a = {\bf x}_a - {\bf x}$ is the vector of relative
separation between the particle of spin $\mbox{\boldmath$\sigma$}$ at ${\bf x}$ and the center of mass of the source $a$. 
Equation (25) can be expressed as $H=c(\mbox{\boldmath$\sigma$}\cdot\partial_{\bf x})\Phi_g$, where

\begin{equation}
\Phi_g = \frac{G}{c^3}\sum_a\;\;\frac{{\bf J}_a\cdot{\bf r}_a}{r^3_a}
\end{equation}

\noindent is the net dimensionless scalar gravitomagnetic potential defined by ${\bf B}_{g} =
c^2\mbox{\boldmath$\nabla$}\Phi_g$.  For a laboratory experiment near the Earth, it is simple to show that the net
contribution due to the Sun, the Moon and the other planets is negligible.  Therefore, to compute $\Phi_g$ one must
investigate the cosmic mass-current distribution.  This is a difficult observational problem and much remains unknown
regarding the distribution of angular momentum in the universe.  It is likely that over the largest scales no preferred
sense of rotation would be discernible.  These considerations lead one to surmise that near the Earth (or Jupiter) the main
contribution to the Hamiltonian is simply due to the Earth (or Jupiter), though a completely satisfactory resolution is
not available.  Conversely, observational data regarding the gravitomagnetic spin-rotation coupling could in principle set
limits on the cosmic mass-current distribution.

\vspace{.25in}
\noindent{\large 4 Discussion}\\

The gravitational coupling of intrinsic spin with rotation has been described in this work and the consequences of the
gravitomagnetic interaction $H=\mbox{\boldmath$\sigma$}\cdot\mbox{\boldmath$\Omega$}_P$ have been pointed out.  In particular, the
gravitomagnetic shift in the Larmor frequency of a nuclear particle has been estimated.  Efforts are under way to improve
the sensitivity of measurement of such frequency shifts by several orders of magnitude.  This could potentially make the
effect measurable near the surface of Jupiter [30].  Let us recall that for Jupiter $\hbar\Omega_{J}\simeq 10^{-27} {\rm
eV}$, corresponding to a gravitomagnetic Larmor shift of about $3 \times 10^{-13}{\rm Hz}$.  In view of the current interest in
planetary exploration, it appears that the gravitomagnetic coupling of intrinsic spin with rotation could be measurable in the
foreseeable future. 

\vspace{.25in}
\noindent {\large Acknowledgements}\\

I am grateful to Friedrich Hehl and Michael Romalis for helpful discussions and correspondence.

\newpage
\noindent {\large References}
\begin{description}
\item{[1]}  Kobzarev I Yu and Okun LB 1963 {\it JETP} {\bf 16} 1343
\item{[2]}  Leitner J and Okubo S 1964 {\it Phys. Rev. B} {\bf 136} 1542\\ Hari Dass ND 1976 {\it Phys. Rev. Lett.} {\bf
36} 393\\ Peres A 1978 {\it Phys. Rev. D} {\bf 18} 2739\\ Morgan TA and Peres A 1962 {\it Phys. Rev. Lett.} {\bf 9} 79
\item{[3]}  Dabbs JWT, Harvey JA, Paya D and Horstmann H 1965 {\it Phys. Rev. B} {\bf 139} 756
\item{[4]}  Velyukhov GE 1968 {\it JETP Lett.} {\bf 8} 229
\item{[5]} Vasil'ev BV 1969 {\it JETP Lett.} {\bf 9} 175
\item{[6]} Young BA 1969 {\it Phys. Rev. Lett.} {\bf 22} 1445
\item{[7]} Wineland DJ and Ramsey NF 1972 {\it Phys. Rev. A} {\bf 5} 821
\item{[8]} Hayasaka H and Takeuchi S 1989 {\it Phys. Rev. Lett.} {\bf 63} 2701
\item{[9]} Faller JE, Hollander WJ, Nelson PG and McHugh MP 1990 {\it Phys. Rev. Lett.} {\bf 64} 825\\
Quinn TJ and Picard A 1990 {\it Nature} {\bf 343} 732\\
Nitschke JM and Wilmarth PA 1990 {\it Phys. Rev. Lett.} {\bf 64} 2115
\item{[10]} Adelberger EG, Heckel BR, Stubbs CW and Rogers WF 1991 {\it Ann. Rev. Nucl. Part. Sci.} {\bf 41} 269
\item{[11]} Wineland DJ {\it et al.} 1991 {\it Phys. Rev. Lett.} {\bf 67} 1735
\item{[12]} Venema BJ {\it et al.} 1992 {\it Phys. Rev. Lett.} {\bf 68} 135
\item{[13]} Ritter RC, Winkler LI and Gillies GT 1993 {\it Phys. Rev. Lett.} {\bf 70} 701
\item{[14]} Berglund CJ {\it et al.} 1995 {\it Phys. Rev. Lett.} {\bf 75} 1879; Youdin AN {\it et al.} 1996 {\it Phys. Rev. Lett.}
{\bf 77} 2170
\item{[15]} Ni W-T {\it et al.} 1999 {\it Phys. Rev. Lett.} {\bf 82} 2439\\ Chui TCP and Ni W-T 1993 {\it Phys. Rev. Lett.} {\bf 71}
3247
\item{[16]} Vorobyov PV and Gitarts Ya I 1988 {\it Phys. Lett. B} {\bf 208} 146\\ Bobrakov VF {\it et al.} 1991 {\it JETP Lett.}
{\bf 53} 294\\ Vorob'ev PV 1994 {\it JETP Lett.} {\bf 59} 510
\item{[17]} Jacobs JP {\it et al.} 1995 {\it Phys. Rev. A} {\bf 52} 3521
\item{[18]} Werner SA, Staudenmann J-L and Colella R 1979 {\it Phys. Rev. Lett.} {\bf 42} 1103\\ Rauch H and Werner SA 2000 {\it
Neutron Interferometry} (Clarendon Press, Oxford)\\ Atwood DK, Horne MA, Shull CG and Arthur J 1984 {\it Phys. Rev. Lett.} {\bf 52}
1673\\ Bonse U and Wroblewski T 1983 {\it Phys. Rev. Lett.} {\bf 51} 1401\\ Hasselbach F and Nicklaus M 1993 {\it Phys. Rev. A} {\bf
48} 143\\ Moorhead GF and Opat GI 1996 {\it Class. Quantum Grav.} {\bf 13} 3129\\ Schwab K, Bruckner N and Packard RE 1997 {\it
Nature} {\bf 386} 585\\ Gustavson TL, Bouyer P and Kasevich MA 1997 {\it Phys. Rev. Lett.} {\bf 78} 2046
\item{[19]} Mashhoon B 1995 {\it Phys. Lett. A} {\bf 198} 9\\
Mashhoon B, Neutze R, Hannam M and Stedman GE 1998 {\it Phys. Lett. A} {\bf 249} 161
\item{[20]} Beth RA 1936 {\it Phys. Rev.} {\bf 50} 115
\item{[21]} Mashhoon B 1988 {\it Phys. Rev. Lett.} {\bf 61} 2639; 1992 {\it ibid.} {\bf 68} 3812\\
Mashhoon B 1989 {\it Phys. Lett. A} {\bf 139} 103\\
Mashhoon B 1990 {\it Phys. Lett. A} {\bf 143} 176\\
Mashhoon B 1990 {\it Phys. Lett. A} {\bf 145} 147\\
Mashhoon B 1993 {\it Phys. Rev. A} {\bf 47} 4498
\item{[22]} Mashhoon B 1993 {\it Phys. Lett. A} {\bf 173} 347\\
Mashhoon B, Gronwald F and Theiss DS 1999 {\it Ann. Physik} {\bf 8} 135\\
Mashhoon B, Gronwald F and Lichtenegger HIM 2000 in {\it Testing Relativistic Gravity in Space}, edited by C. L\"{a}mmerzahl, C.W.F.
Everitt and F.W. Hehl (Springer-Verlag, Berlin); gr-qc/9912027
\item{[23]} de Oliveira CG and Tiomno J 1962 {\it Nuovo Cimento} {\bf 24} 672\\
Mitskievich NV 1969 {\it Physical Fields in General Relativity Theory}, in Russian (Nauka, Moscow)\\
Schmutzer E 1973 {\it Ann. Physik} {\bf 29} 75\\
Barker BM and O'Connell RF 1975 {\it Phys. Rev. D} {\bf 12} 329\\
Schmutzer E and Pleba\'{n}ski J 1977 {\it Fortschr. Phys.} {\bf 25} 37
\item{[24]} Hehl FW and Ni W-T 1990 {\it Phys. Rev. D} {\bf 42} 2045\\
Cai YQ and Papini G 1991 {\it Phys. Rev. Lett.} {\bf 66} 1259; 1992 {\it ibid.} {\bf 68} 3811\\
Anandan J 1992 {\it Phys. Rev. Lett.} {\bf 68} 3809\\
Andretsch J and L\"{a}mmerzahl C 1992 {\it Appl. Phys. B} {\bf 54} 351\\
Silverman MP 1992 {\it Nuovo Cimento D} {\bf 14} 857\\
Huang J 1994 {\it Ann. Physik} {\bf 3} 53\\
Soares ID and Tiomno J 1996 {\it Phys. Rev. D} {\bf 54} 2808\\
Ryder LH 1999 {\it Gen. Rel. Grav.} {\bf 31} 775
\item{[25]} Mashhoon B 1974 {\it Nature} {\bf 250} 316\\
Mashhoon B 1974 {\it Phys. Rev. D} {\bf 10} 1059\\
Mashhoon B 1975 {\it Phys. Rev. D} {\bf 11} 2679
\item{[26]} Damour T and Ruffini R 1974 {\it C.R. Acad. Sci. A} {\bf 279} 971\\
de Logi WK and Kovacs SJ Jr. 1977 {\it Phys. Rev. D} {\bf 16} 237\\
Leahy DA 1982 {\it Int. J. Theor. Phys.} {\bf 21} 703\\
Mashhoon B 1987 {\it Phys. Lett. A} {\bf 122} 299\\
Futterman JAH, Handler FA and Matzner RA 1988 {\it Scattering from Black Holes} (Cambridge University Press, Cambridge)\\
Feng LL and Lu T 1991 {\it Class. Quantum Grav.} {\bf 8} 851\\
Carini P, Feng LL, Li M and Ruffini R 1992 {\it Phys. Rev. D} {\bf 46} 5407
\item{[27]} Harwit M {\it et al.} 1974 {\it Nature} {\bf 249} 230\\
Dennison B, Dickey J and Jauncey D 1976 {\it Nature} {\bf 263} 666\\
Dennison B {\it et al.} 1978 {\it Nature} {\bf 273} 33
\item{[28]} Mashhoon B 1999 {\it Gen. Rel. Grav.} {\bf 31} 681
\item{[29]} Kopeikin S and Mashhoon B 2000 preprint
\item{[30]} Romalis M 2000 private communication
\end{description} 

\end{document}